\documentclass[twocolumn]{jpsj2} 
\usepackage{amsmath}

\title{
Specific Heat-Coefficient of YbAl$_3$ Studied by Combined Nearly Free Electron Conduction Band Hybridized with Localized f Electrons with Correlation Effect
}

\author{%
Yoshiki {\sc Imai} and Tetsuro {\sc Saso}
}

\inst{%
Department of Physics, Saitama University, Saitama, 338-8570
}

\recdate{\today}

\abst{
Based on the recently proposed band model, the electronic specific heat of moderately heavy electron compound YbAl$_3$ are investigated. The band term of the Hamiltonian consists of three parts; conduction electrons described by the nearly free electron method, localized 4f electrons of Yb ions and the hybridization term between these electrons. Extracting several bands near the Fermi level, we reconstruct the low-energy effective Hamiltonian in order to consider the correlation effect, which is studied by using the self-consistent second order perturbation theory combined with local approximation. 
The temperature dependence of the specific heat $c_{\rm v}(T)$ is calculated as a function of temperature $T$ from the numerical derivative of the internal energy. 
Sommerfeld coefficient $\gamma$ is also calculated from the direct formula. 
The overall structure of $c_{\rm v}(T)/T$ is in quantitative agreement with the experimental results, 
which have the characteristic two-peak structures. They originate from the correlation effect and the structure of the non-interacting density of states, respectively. 
We show that our effective Hamiltonian yielding the realistic band structure may describe quantitatively heavy electron compounds with conduction bands composed of s- or p- electrons. 
}

\kword{
YbAl$_3$, nearly free electron model, empty-core pseudo-potential
}
\begin{document}
\maketitle
%
\section{Introduction}
The periodic Anderson model (PAM) and its extensions are often employed in order to investigate the physical properties of heavy fermion compounds.\cite{hews93} These models succeeded in capturing the qualitative properties. However, much simplification is done in PAM, sometimes leading to incorrect results even qualitatively, and it is difficult to describe quantitative properties of realistic materials. Therefore, tight-binding model is constructed form the band calculations based on the local density approximation (LDA) and correlation effects are incorporated in various ways. 

Recently we proposed a new method of the construction of the effective Hamiltonian with the realistic band structure for heavy fermion materials whose conduction bands consist of s- and p- electrons mainly.\cite{kuro07}
By using the combination of solid methods, we tried to construct the quantitatively reliable model with only a few parameters.
It is well known that the band calculation for heavy fermion compounds sometimes reproduces incorrect f electron level. 
Thus, we can study the realistic materials from another point of view by means of this technique. Furthermore, this model can capture the basic properties of the realistic materials such as the lattice symmetry, and the various physical quantities can be calculated quantitatively from almost absolute zero to finite temperature region within the desired accuracy in the wave-vector $\mathbf k$ and frequency $\omega$ spaces. Thus, this model may be good starting point to discuss the physical properties of strongly correlated systems. 

The Hamiltonian consists of conduction electrons described by the nearly free electron (NFE) method, localized f electrons and the hybridization between the conduction and the localized f electrons. This method includes only a few parameters. We apply it to typical heavy fermion compound YbAl$_3$, which has NFE-like conduction bands below and around the Fermi energy $E_{\rm F}$. 
The obtained band structure is in good agreement with the result of the band calculation provided by Harima \cite{kuro07,hari}, and the optical conductivity is also consistent with the experimental result.\cite{okam07} 
However, the calculated value of Sommerfeld coefficient $\gamma$ is about 13.5 mJ$\cdot$K$^{-2}\cdot$mol$^{-1}$,\cite{corn02,ebih00} which is 1/3 of the experimental result 40 mJ$\cdot$K$^{-2}\cdot$mol$^{-1}$. \cite{corn02,baue04}
This discrepancy may result from the lack of the correlation effect 
which was not taken into account in the previous study. 

Note here the physical properties for YbAl$_3$. 
The mean valence is in the range from 2.65 to 2.8\cite{tjen93,suga05} and the Sommerfeld coefficient $\gamma$ is about 40 mJ$\cdot$ K$^{-2}$$\cdot$ mol$^{-1}$,  
which indicates this material is a valence fluctuation compound and does not have strong correlation effect. The amplitude of the crystal field splitting of 4f $J=7/2$ level is below 10 meV\cite{walt91}, which is much smaller than the Kondo temperature $T_{\rm K}\sim$ 500 K\cite{corn02}.  
In the low-temperature region, the specific heat divided by temperature $c_{\rm v}(T)/T$ shows prominent temperature dependence; a peak structure appears at $T\sim$ 70 K where the amplitude is about 80 mJ$\cdot$ K$^{-2}$$\cdot$ mol$^{-1}$.
Furthermore, $c_{\rm v}(T)/T$ decreases from $T=0$ below $T\sim 25$ K, which was interpreted as the effect of the lattice coherence \cite{corn02}. 
The coherence temperature is about $T_{\rm coh} \sim 35$ K, below which the electrical resistivity is proportional to $T^2$ \cite{corn02,ebih03}. This result indicates that YbAl$_3$ satisfies the Fermi liquid behavior. The optical conductivity shows the mid-infrared peak structure at $\hbar \omega \sim 0.25$ eV.\cite{okam04,okam07} 

In this paper, we focus on the heavy fermion metal YbAl$_3$ and construct the low-energy effective Hamiltonian with the realistic band structure and the Coulomb interaction. The latter term is not included in the previously proposed effective Hamiltonian. 
We investigate the physical properties and discuss the efficiency of the Hamiltonian. 
Applying the self-consistent second order perturbation theory with local approximation to this low-energy effective Hamiltonian, we calculate $c_{\rm v}(T)/T$ and the Sommerfeld coefficient. 

This paper is organized as follow. We review the band model for YbAl$_3$ with the realistic band structure and construct the low-energy effective Hamiltonian with the Coulomb interaction to discuss the correlation effect in the next section. In \S3, we show the density of states (DOS), the Sommerfeld coefficient and the specific heat divided by temperature. Summary and discussions are given by in \S4.

%
\section{Effective Hamiltonian for YbAl$_3$ and Method}
In this section, the low-energy effective Hamiltonian for YbAl$_3$ is constructed, which can be written as follows 
\begin{eqnarray}
H^{\rm eff}&=&H^{\rm band}+H^{\rm int},\\
H^{\rm band}&=&H^{\rm c}+H^{\rm f}+H^{\rm hyb},
\end{eqnarray}
where $H^{\rm band}$ ($H^{\rm int}$) represents the band (interaction) part, respectively. 
$H^{\rm band}$ consists of the conduction band $H^{\rm c}$, the localized 4f orbitals $H^{\rm f}$, and the hybridization between the conduction and the localized 4f-electrons term $H^{\rm hyb}$, while $H^{\rm int}$ consists of the Coulomb interaction between the localized 4f-electrons of Yb ions. 

We briefly review the band part of the Hamiltonian $H^{\rm band}$ based on the proposed method\cite{kuro07}. Each electronic configuration of YbAl$_3$ is  [Ne](3s)$^2$(3p)$^1$ for Al and [Xe](4f)$^{14}$(5d)$^0$(6s)$^2$ for Yb atom. There are 25 electrons per unit cell of YbAl$_3$ except the closed shells. The mean configuration of the localized 4f-electrons lies between f$^{14}$ and f$^{13}$, which indicates that YbAl$_3$ shows the valence fluctuation behavior and becomes a typical metallic system. Most of the conduction electrons consist of s- and p-electrons, which wholly spread over the crystal with the weak periodic potentials. Therefore, for the conduction bands, we employ the NFE method with the empty-core pseudo-potential, whose screening effect is considered by using the dielectric function of the free electron gas. 
The atomic potential and the core radius in each atom are employed from Harrison's textbook \cite{harr80,harr99}. 

The 14-fold degenerated 4f-electron levels split into the total angular momentum $J$=$7/2$ (8-fold degeneracy) and $J$=$5/2$ (6-fold degeneracy) levels because of the strong spin-orbit coupling in the case of YbAl$_3$. The difference between both energy levels is of the order of 0.1 Ry, which is obtained from the band calculation \cite{hari}. The $J$=$7/2$ level $E_{\rm f}^{7/2}$ is located near and below the Fermi level, while the $J$=$5/2$ level is far from the Fermi level \cite{ebih00}. 
Although the $J$=$5/2$ level may not affect the physical properties in the low-temperature region, we also include $J$=$5/2$ term in the Hamiltonian $H^{\rm band}$ and set the energy difference between both levels as 0.1 Ry. Thus $E_{\rm f}^{7/2}$ is a tuning parameter in this method. 
Note here that the amplitude of the crystal field splitting of 4f $J=7/2$ level is very small compared to the Kondo temperature. 
Therefore, although to consider the crystal field splitting effect is easy in this framework, it is not discussed in this paper. 

The plane wave with wave vector $\mathbf k$ and the localized 4f-electron can be expanded by the spherical harmonics, so that the matrix element of the hybridization term is also described by this spherical harmonics. 
For simplicity, we neglect {\bf k}-dependence of the radial part and regard this part as a constant which denotes the hybridization amplitude $V$. 
We stress that although we neglect the {\bf k}-dependence of the radial part, we keep the angular dependence of {\bf k} completely. 

\begin{figure}[tb]
\begin{center}
\includegraphics[width=8.5cm]{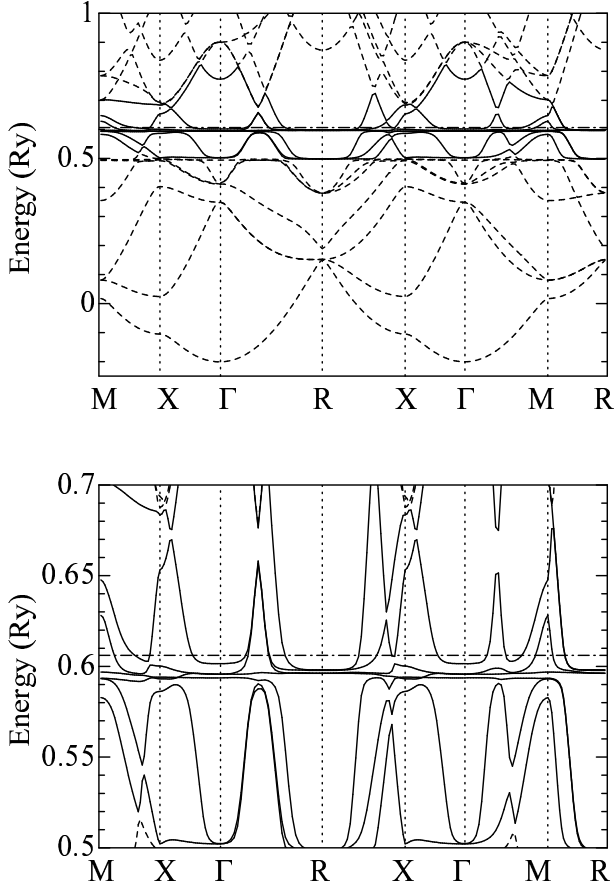}
\end{center}
\caption{Energy band dispersion $E_{\mathbf k \alpha}$ of YbAl$_3$. The 4f $J=7/2$ level is set at $E_{\rm f}^{7/2}=0.6$ Ry. The solid lines stand for the extracted bands in order to consider the correlation effect. The dashed lines are the bands not included in the effective Hamiltonian eq. (\ref{eqn:effham}).
The dot-dashed line stands for the Fermi level. 
The lower panel is the magnification of the upper panel near the Fermi level. 

}
\label{band_ext}
\end{figure}

The constructed band term of the Hamiltonian is given by
\begin{eqnarray}
H^{\rm band}=
\left (
\begin{array}{cc|c}
H^{\rm c}_{\uparrow} 
& 0 
&\vspace{-2mm}
\\

&
&{\cal H}^{\rm hyb} 
\\

0 
& {\cal H}^{\rm c}_{\downarrow} 
&  \\ \hline

{\cal H}^{\rm hyb*}\hspace{-10mm}
& 
& {\cal H}^{\rm f}
\end{array}
\right). 
\end{eqnarray}
Diagonalizing the band term of the Hamiltonian, we can rewrite as 
\begin{eqnarray}
H^{\rm band}=\sum_{\mathbf k \alpha}E_{\mathbf k \alpha} a^{\dag}_{\mathbf k \alpha} a_{\mathbf k \alpha}, 
\end{eqnarray}
where $E_{\mathbf k \alpha}$ denotes an energy eigenvalue, and $a_{\mathbf k \alpha}$ is an annihilation operator of a quasi-particle with the wave-vector ${\mathbf k}$ and the band index $\alpha$. The energy bands are shown in Fig.\ref{band_ext}, which is in rough agreement with the LDA result around and below the Fermi level except $M$-point. \cite{hari}
The topology of the Fermi surface is also consistent with the LDA result except the same point. 

Next, let us consider the interaction term $H^{\rm int}$. 
The correlation effect may be essential on the localized 4f-electrons at the low-temperature region, so we assume that $H^{\rm int}$ is given by 
\begin{eqnarray}
H^{\rm int}=U\sum_{i, M<M',J=\frac{7}{2}}n_{iJM}n_{iJM'},
\end{eqnarray}
where $n_{iJM}$ is a number operator of the 4f electron with $J$. $M$ is the $z$-component of $J$ and $U$ stands for the Coulomb interaction. 

By using the diagonalized bases for the band term, an annihilation operator of the 4f electron with the wave-vector ${\mathbf k}$ and the angular momentum $J$ and its $z$-component $M$ is written as
\begin{eqnarray}
f_{\mathbf k JM}=\sum_{\alpha} u_{\mathbf k J M \alpha} a_{\mathbf k \alpha},
\end{eqnarray}
where $u_{\mathbf k J M \alpha}$ stands for the eigenvector of $H^{\rm band}$. 
Thus, $H^{\rm int}$ is written as
\begin{eqnarray}
H^{\rm int}=\frac{1}{N}\sum_{\mathbf k,\mathbf k',\mathbf q}\sum_{\alpha \alpha' \beta \beta'}U_{\mathbf k,\mathbf k',\mathbf q}^{\alpha \alpha' \beta \beta'}a^{\dag}_{\mathbf{k+q} \alpha}a_{\mathbf{k} \alpha'}a^{\dag}_{\mathbf{k'-q} \beta}a_{\mathbf{k'} \beta'}, 
\end{eqnarray}
where $N$ is the number of the lattice sites, and the interaction matrix $U_{\mathbf k,\mathbf k',\mathbf q}^{\alpha \alpha' \beta \beta'}$ is as follows 
\begin{eqnarray}
U_{\mathbf k,\mathbf k',\mathbf q}^{\alpha \alpha' \beta \beta'}=
U\sum_{M<M'}u_{\mathbf{k+q}JM \alpha}u^{*}_{\mathbf{k} JM\alpha'}u^{*}_{\mathbf{k'-q} JM'\beta}u_{\mathbf{k'} JM'\beta'}. 
\end{eqnarray}

The hybridization does not affect the band dispersions drastically. 
There are almost flat bands near and below the Fermi level, which indicates that these diagonalized bands still hold the 4f character strongly (shown in Fig.\ref{band_ext}). 
Thus we extract 14 bands near the Fermi level, which almost lie in $|E_{\mathbf k \alpha} -E_{\rm F}| < 0.1$ Ry and include all the 4f $J=7/2$ components. 
There are sufficient to discuss the effect of the Coulomb interaction. 

To avoid the extremely complicated treatment, we regard the interaction matrix as diagonal within the extracted band indices as in eq. (\ref{eqn:hubint}) below. 
Then, the interaction term can be expressed in the Hubbard type with the number operators of the extracted bands, $n_{i\alpha}n_{i\beta}$. 
In addition, the amplitude is approximated by a constant. 
Thus the interaction matrix is written as
\begin{eqnarray}
U_{\mathbf k,\mathbf k',\mathbf q}^{\alpha \alpha' \beta \beta'} \sim \tilde{U}
\delta_{\alpha \alpha'} \delta_{\beta \beta'}.
\label{eqn:hubint} 
\end{eqnarray}
This treatment means that we allow Coulomb interaction to act both on the 4f and conduction electrons equally. However, it acts on the 4f electrons more strongly since the DOS is much larger. 
Hereafter, we regard the effective Coulomb interaction $\tilde{U}$ as a parameter to discuss the correlation effect. 
Since Coulomb interaction is strongly screened and renormalized in metallic heavy fermion compounds, the amplitude may not be so large at the low-energy region. 

Note that the band index is labeled in the order of energy eigenvalue from smallest to largest at each wave-vector $\mathbf k$, for simplicity. 
One should classify the band index according to the space symmetry, but the simplification for the labeling is irrelevant to the physical quantities such as specific heat in the present study. 

Finally, we obtain the following low-energy effective Hamiltonian
\begin{eqnarray}
H^{\rm eff}&=&\sum_{\mathbf k \alpha}E_{\mathbf k \alpha} a^{\dag}_{\mathbf k \alpha} a_{\mathbf k \alpha}\nonumber \\
&+&\frac{\tilde{U}}{N}\sum_{\mathbf k,\mathbf k',\mathbf q}\sum_{\alpha < \beta}a^{\dag}_{\mathbf{k+q} \alpha}a_{\mathbf{k} \alpha}a^{\dag}_{\mathbf{k'-q} \beta}a_{\mathbf{k'} \beta}. 
\label{eqn:effham}
\end{eqnarray}
Here, $E_{\mathbf k \alpha}$ consists of only the extracted bands near the Fermi level. 

Let us consider the correlation effect based on this low-energy effective Hamiltonian with the realistic band structure. The amplitude of the Coulomb interaction is not so large because of the renormalization effect at the low-energy region, so that we employ the self-consistent second order perturbation theory with respect to the Coulomb interaction. Furthermore, since the 4f electrons have strongly localized character, local approximation is applied. Note that the off-diagonal components in the self energy vanish in the present approximation since the Hubbard-type interaction is employed. 

The Green function and the local Green function are defined as 
\begin{eqnarray}
G_{\alpha}(\mathbf k, {\rm i} \omega_n) 
&=&\left[ {\rm i} \omega_n + \mu -E_{\mathbf k \alpha} -\Sigma_{\alpha} ({\rm i}\omega_n)\right]^{-1}, \\
G^{\rm loc}_{\alpha}({\rm i} \omega_n) 
&=&\frac{1}{N}\sum_{\mathbf k}G_{\alpha}(\mathbf k, {\rm i} \omega_n), 
\end{eqnarray}
where $\omega_n$ stands for Matsubara frequency and $\mu$ is the chemical potential. Combined with local approximation, the self energy is given by
\begin{eqnarray}
\Sigma_{\alpha}({\rm i}\omega_n)
&=&\tilde{U}T\sum_{n',\alpha'(\ne \alpha)}G^{\rm loc}_{\alpha'}({\rm i} \omega_{n'}) 
e^{{\rm i} \omega_{n'} 0^+} \nonumber \\
&&-\,\tilde{U}^2T^2\sum_{m,m', \alpha'(\ne \alpha)}G^{\rm loc}_{\alpha}({\rm i} \omega_n-{\rm i} \omega_m)\nonumber \\
&&\times G^{\rm loc}_{\alpha'}({\rm i} \omega_m+{\rm i} \omega_{m'})G^{\rm loc}_{\alpha'}({\rm i} \omega_{m'}), 
\end{eqnarray}
where $0^+$ is a positive infinitesimal number. $T$ denotes temperature which includes the Boltzmann constant and has a dimension of energy. 
The internal energy is written as 
\begin{eqnarray}
E^{\rm int}&=&\sum_{\mathbf k \alpha}E_{\mathbf k \alpha} 
\left\{ 
T
\sum_{n}G_{\alpha}(\mathbf k, {\rm i} \omega_n) 
e^{{\rm i} \omega_{n} 0^+} 
\right\}\nonumber \\
&+&
\frac{NT}{2} 
\sum_{n, \alpha}G^{\rm loc}_{\alpha}({\rm i} \omega_n) \Sigma_{\alpha}({\rm i} \omega_n). 
\end{eqnarray}
The sum of Matsubara frequency is replaced by the integration of real frequency by means of analytic continuation. 

The specific heat is obtained from the derivative of the internal energy with respect to temperature, which is given by 
\begin{eqnarray}
c_{\rm v}(T)=\frac{N_{\rm A}}{N}\frac{{\rm d}E^{\rm int}}{{\rm d}T}, 
\label{eqn:cv}
\end{eqnarray}
where $N_{\rm A}$ denotes Avogadro constant. 
The estimation of the derivative is replaced actually with the numerical difference. 

The Sommerfeld coefficient $\gamma$ is given by 
\begin{eqnarray}
&&\gamma =\frac{\pi^2 k_{\rm B}^2 N_{\rm A}}{3N} \sum_{\mathbf k \alpha} 
A_{\alpha}(\mathbf k,0)
\left\{1-\frac{\partial \Sigma_{\alpha}(\omega)}{\partial \omega}\bigg|_{\omega =0} \right\}, \\
&&A_{\alpha}(\mathbf k,\omega)=
-\frac{1}{\pi} {\rm Im}\,G_{\alpha}(\mathbf k, \omega), 
\end{eqnarray}
within the Fermi liquid theory with local approximation. $A_{\alpha}(\mathbf k,0)$ stands for the spectral function on the Fermi energy. 
For metallic systems with the Fermi liquid behavior, $\lim_{T\rightarrow 0}c_{\rm v}(T)/T$ must be equal to $\gamma$. Note that YbAl$_3$ exhibits the Fermi liquid property at the low temperature region ($T<35$ K). 

The total DOS and the number of 4f-electrons with the angular momentum $J$ are given by 
\begin{eqnarray}
\rho(\omega)&=&\sum_{\mathbf k \alpha} A_{\alpha}(\mathbf k,\omega),\\
n_{\rm f}^{J}&=&\frac{1}{N}\sum_{\mathbf k, M,\alpha}|u_{\mathbf k JM \alpha}|^2
\int \, {\rm d}\omega\,
A_{\alpha}(\mathbf k,\omega) f(\omega), 
\end{eqnarray}
where $f(\omega)$ stands for Fermi distribution function. 

In the present method, the Hamiltonian includes only a few parameters (4f $J=7/2$ level $E_{\rm f}^{7/2}$, the hybridization amplitude $V$ and the effective Coulomb interaction $\tilde{U}$). 
Compared with some experimental results, these parameters are estimated. 

\section{Results}
In the NFE method, the weak periodic potential is given by the sum of reciprocal lattice vectors as 
\begin{eqnarray}
V(\mathbf r)&=&\sum_{l} V_{\mathbf G_l} {\rm e}^{{\rm i} \mathbf G_l \cdot \mathbf r}, \\
\mathbf G_l &\equiv&l_1 \mathbf G_0^1+l_2 \mathbf G_0^2+l_3 \mathbf G_0^3. 
\end{eqnarray}
$\mathbf G_0^{\mu}$ ($\mu=1,2,3$) stands for the primitive translation vector of the reciprocal lattice and $l$ denotes an integer set ($\equiv (l_1,l_2,l_3)$) within our target energy range. 
The calculation, however, converges within $|l_\mu| \leq 3$ ($\mu =1,2,3$). 
We divide the first Brillouin zone into $80 \times 80 \times 80$ pieces to carry out wave vector $\mathbf k$ sums. 
In order to focus on the correlation effect, the tuning parameters for the band structure are given by the same as those in the previous paper.\cite{kuro07} 
Namely, the 4f $J=7/2$ level and the hybridization amplitude are fixed mainly as $E_{\rm f}^{7/2}=0.6$ Ry and $V=0.03$ Ry, respectively. 
The 4f $J=5/2$ level is defined as $E_{\rm f}^{5/2}=E_{\rm f}^{7/2}-0.1$ Ry. The pseudo-potential parameters are defined as follows; core radius  $r_{\rm c}^{\rm Yb}=2.0$ \AA\, for Yb and $r_{\rm c}^{\rm Al}=0.61$ \AA\, for Al \cite{harr80,harr99}. Note that a larger value for $r_{\rm c}^{\rm Yb}$ than that of ref. \ref{harr80} is chosen.\cite{kuro07} The valence of each atom is $Z^{\rm Yb}=+2$ and $Z^{\rm Al}=+3$, respectively.  

\begin{figure}[bt]
\begin{center}
\includegraphics[width=8.5cm]{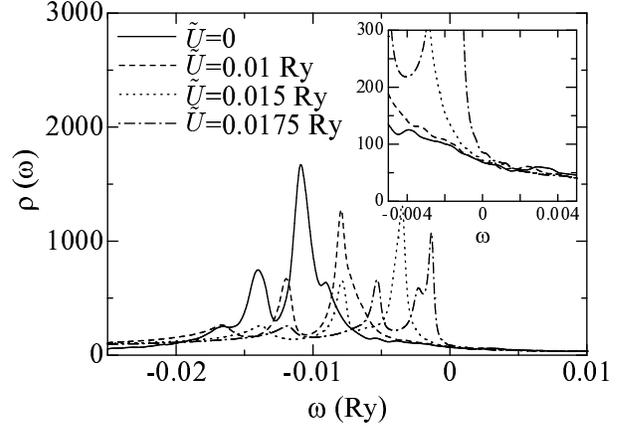}
\end{center}
\caption{DOS as a function of the effective Coulomb interaction $\tilde{U}$ for $E_{\rm f}^{7/2}=0.6$ Ry at $T=2.0\times 10^{-5}$ Ry. The Fermi level is set to $\omega=0$. The inset shows the magnification near the Fermi level. 
}
\label{dos}
\end{figure}

Figure \ref{dos} shows the total DOS for a few choices of $\tilde{U}$ near the low-energy region and at the low-temperature region. 
The prominent peak structures appear below the Fermi level ($\omega=0$), which mainly consist of the 4f $J=7/2$ electrons. Although the spectrum is broadened due to the hybridization effect, the localized character of 4f electrons still remains.

The simple PAM or its extension produces a hybridization gap easily even in the metallic system, which is caused by the simplification, such as neglect of the wave vector dependence in the hybridization term. 
In the present model, the hybridization still keeps the angular dependence of $\mathbf k$, so that a hybridization gap never appears. Figure \ref{dos} shows only a pseudo gap structure. 
Note that the Coulomb interaction does not affect topology of the Fermi surface within the present parameter region.

\begin{figure}[b]
\begin{center}
\includegraphics[width=8.5cm]{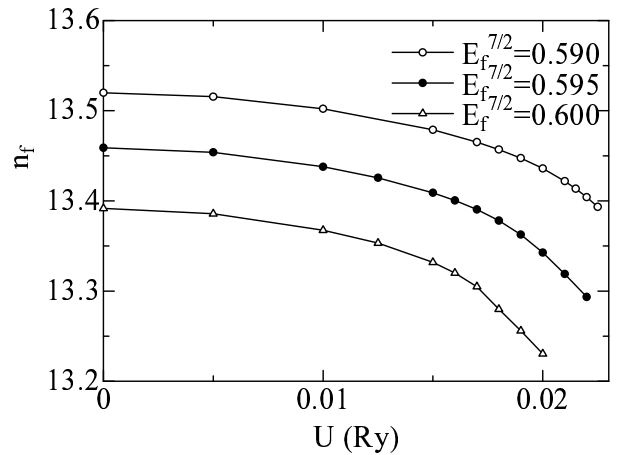}
\end{center}
\caption{The number of 4f electrons as a function of the effective Coulomb interaction $\tilde{U}$ for a few choices of $E_{\rm f}^{7/2}$ at $T=2.0 \times 10^{-5}$ Ry. 
}
\label{nf}
\end{figure}

With an increase of the effective Coulomb interaction, the peak structure is broadened and is shifted to the Fermi level gradually, which means that the effective 4f $J=7/2$ level is shifted by $\tilde{U}$ while the amplitude at $\omega \sim 0$ seems almost insensitive (Fig.\ref{dos} inset). 
Although the number of 4f electrons decreases due to this shift, the deviation from $\tilde{U}=0$ Ry is not serious, which is shown in Fig. \ref{nf} 
for $E_{\rm f}^{7/2}=0.6$ Ry (and other parameters for reference). 
The difference between the case of $\tilde{U}=0$ Ry and that of $\tilde{U}\sim 0.0175$ Ry is less than 0.1, which still stays in the following experimental result. 
The mean valence of Yb ion in YbAl$_3$ is 2.65 $\sim$ 2.8, which indicates the number of 4f electrons is the range from 13.2 to 13.35. 
Of course, further increase of $\tilde{U}$ pushes up the effective 4f level above $E_{\rm F}$, then $n_{\rm f}$ decreases quickly. 

The effective Coulomb interaction is introduced for the diagonalized bands which partially include the conduction bands within the present treatment. 
If we treat the correlation effect only between 4f electrons, the number of 4f electrons may be rather reduced in comparison with the present result. 
However, the weight of the 4f electrons is over 2/3 in the total DOS obtained from the extracted bands, 
so that we believe $n_{\rm f}$ vs. $\tilde{U}$ relation may not change drastically. 

\begin{figure}[bt]
\begin{center}
\includegraphics[width=8.5cm]{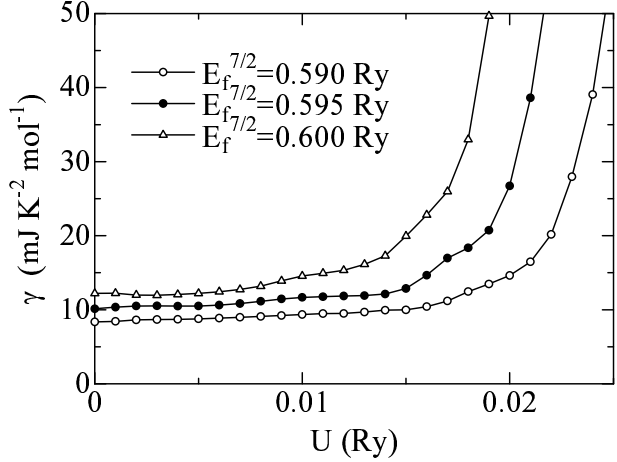}
\end{center}
\caption{The Sommerfeld coefficient as a function of the effective Coulomb interaction $\tilde{U}$ for a few choices of $E_{\rm f}^{7/2}$ at $T=2.0 \times 10^{-5}$ Ry. 
}
\label{gamma}
\end{figure}

Figure \ref{gamma} shows the Sommerfeld coefficient for a few choices of $E_{\rm f}^{7/2}$ at the very low-temperature region. 
Note that all the obtained coefficients are underestimated slightly and systematically since a finite adiabatic constant $\delta$ is introduced in the numerical calculation 
($\rightarrow \omega+ {\rm i}\delta$), 
whose amplitude is given by $\delta=3.0\times 10^{-4}$ Ry. 
For example, $\gamma$ is about 12.5 mJ$\cdot$K$^{-2}$$\cdot$mol$^{-1}$ for $E_{\rm f}^{7/2}=0.6$ Ry for the non-interacting case, whose value is slightly smaller than the result of ref. \ref{kuro07} (13.5 mJ$\cdot$K$^{-2}$$\cdot$mol$^{-1}$). 

With an increase of $\tilde{U}$, $\gamma$ is enhanced and becomes $\sim$ 30 mJ$\cdot$K$^{-2}$$\cdot$mol$^{-1}$ for $\tilde{U}=0.0175$ Ry and $E_{\rm f}^{7/2}=0.6$ Ry. 
This result for the Sommerfeld coefficient is roughly consistent with the experimental one. 
Taking account of the mean valence and the Sommerfeld coefficient observed by the experiments, we can estimate the effective Coulomb interaction as 
$\tilde{U}\sim 0.017 -0.019$ Ry. 

Let us discuss the temperature dependence of the specific heat divided by temperature. 
We estimate eq. (\ref{eqn:cv}) by means of the numerical difference, which is defined as $\tilde{c_{\rm v}}(T)$. 
Although the specific heats $\tilde{c_{\rm v}}(T)$ have to vanish at $T \rightarrow 0$ for any values of $\tilde{U}$, the extrapolation values at $T \rightarrow 0$ have slightly finite amplitude ($\Delta c_{\rm v}$) due to errors on the numerical difference. Thus $\tilde{c_{\rm v}}(T)/T$ is strongly enhanced at $T \rightarrow 0$. 
In order to exclude this artificial anomaly, we subtract $\Delta c_{\rm v}$ from the obtained $\tilde{c_{\rm v}}(T)$ data, where $\Delta c_{\rm v}$ is estimated from $\tilde{c_{\rm v}}(T)$ data at very low-temperatures with least square method. The subtracted specific heat $c_{\rm v}(T)$ is defined as
$
c_{\rm v}(T)\equiv \tilde{c_{\rm v}}(T)-\Delta c_{\rm v}. 
$
All $\Delta c_{\rm v}$ become very small values, which are of the order of $ 1.0 \times 10^{-2}$ mJ$\cdot$K$^{-1} \cdot$ mol$^{-1}$. 
Hereafter we regard $c_{\rm v}(T)$ as the specific heat. 

\begin{figure}[bt]
\begin{center}
\includegraphics[width=8.5cm]{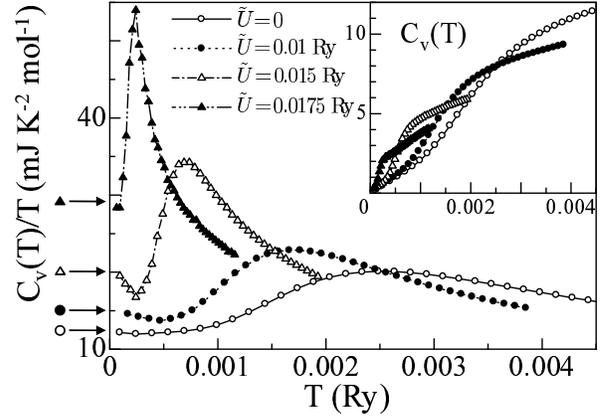}
\end{center}
\caption{$c_{\rm v}(T)/T$ as a function of temperature $T$ for several choices of $\tilde{U}$. The inset shows the temperature dependence of $c_{\rm v}(T)$. The arrows stand for the data of the Sommerfeld coefficients at $T \sim 0$ where each symbol corresponds to that of $c_{\rm v}(T)/T$. 
}
\label{cv}
\end{figure}

\begin{figure}[b]
\begin{center}
\includegraphics[width=8.5cm]{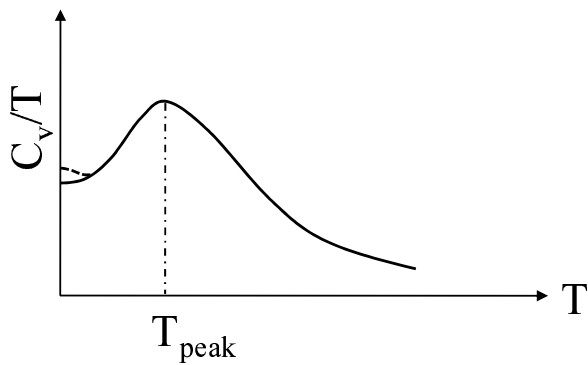}
\end{center}
\caption{ Schematic behavior of $c_{\rm v}(T)/T$ as a function of temperature. 
The solid (dashed) line is the expected $c_{\rm v}(T)/T$ for the non-interacting (interacting) case. 
}
\label{cv2}
\end{figure}

The figure \ref{cv} shows the temperature dependence of $c_{\rm v}(T)/T$, which has to coincide with the Sommerfeld coefficient within the Fermi liquid theory at $T=0$. 
This is approximately achieved in our calculation, as is seen in Fig. \ref{cv}. 
For $\tilde{U}=0$, $c_{\rm v}(T)/T$ is almost independent of temperature at the low temperature region and corresponds to the Sommerfeld coefficient. For the interacting case, $c_{\rm v}(T)/T$ decreases with an increase of temperature from $T=0$ and has the local minimum which is smaller than the Sommerfeld coefficient. 
It seems that $c_{\rm v}(T)/T$ increases with an increase of temperature for the non-interacting case from $T=0$, but decreases for the interacting case. 
With further increase of temperature, $c_{\rm v}(T)/T$ has the peak structure in both cases at $T=T_{\rm peak}$. 
The schematic temperature dependence is shown in Fig. \ref{cv2}. 

Here, we first discuss the behavior of $c_{\rm v}(T)/T$ at $T\sim 0$  for the non-interacting case. 
It is well known that specific heat is proportional to $T$ at low temperature region by means of low temperature expansion of the energy up to $T^2$, so that $c_{\rm v}(T)/T$ is independent of temperature. 
Taking account of the expansion up to $T^4$ for the non-interacting case, the temperature dependences of the chemical potential and $c_{\rm v}(T)/T$ are given by
\begin{eqnarray}
\mu(T)&=&\mu_0-\frac{\pi^2 k_{\rm B}^2}{6}\frac{D^{(1)}}{D^{(0)}}T^2\nonumber \\ 
&-&\frac{\pi^4 k_{\rm B}^4}{36}
\left\{
\frac{7D^{(3)}}{D^{(0)}}
-\frac{D^{(1)}D^{(2)}}{{D^{(0)}}^{2}}
\right\}T^4, \\
\frac{c_{\rm v}(T)}{T}&=&\frac{\pi^2 k_{\rm B}^2}{3}D^{(0)}\nonumber \\
&+&\frac{\pi^4k_{\rm B}^4}{6}
\left\{
\frac{7}{5}D^{(2)}-\frac{{\left(D^{(1)}\right)}^2}{D^{(0)}}
\right\}
T^2,
\label{eqn:cv/t}
\\
D^{(n)}&=&\frac{{\rm d}^{n} D(\omega)}{{\rm d}\omega^n}\bigg|_{\omega=\mu_0},
\end{eqnarray}
where $\mu_0$ stands for the chemical potential at $T=0$ and $D(\omega)$ is the DOS for general systems. \cite{shim81}
For example, when the Lorentzian-like DOS $D(\omega)\propto(\omega^2+a^2)^{-1}$ is assumed, the second term of the right side in eq.(\ref{eqn:cv/t}) is written as
\begin{eqnarray}
\frac{7}{5}D^{(2)}-\frac{{\left(D^{(1)}\right)}^2}{D^{(0)}}
\propto
11\mu_0^2-7a^2, 
\end{eqnarray}
where $a$ ($>$0) is a half width. When $\mu_0$ is larger than $a$ and positive, $D(\omega)$ is almost filled. 
Then, this term becomes positive. 
In this situation, $c_{\rm v}(T)/T$ may increase monotonically with an increase of temperature from $T=0$. 

The present DOS roughly has a Lorentzian-like structure and the large spectrum due to the 4f $J=7/2$ level below the Fermi level. Since the DOS has a tail at $\omega \sim 0$, the similar picture may be realized. Namely, $c_{\rm v}(T)/T$ is enhanced with an increase of temperature. 
Note that the above-mentioned behavior occurs in the low-temperature region due to the low temperature expansion. 

For the interacting case, while it is difficult to obtain the analytical form of $c_{\rm v}(T)/T$ at $T\sim 0$, the obtained result shows that the amplitude of the local minimum in $c_{\rm v}(T)/T$ is smaller than that of each Sommerfeld coefficient and $c_{\rm v}(T)/T$ decreases with an increase of temperature from $T=0$. \cite{imai09} 
In general, with an increase of the Coulomb interaction, the width of the peak in $c_{\rm v}(T)/T$ at $T\sim 0$ becomes narrower since the characteristic energy scale is reduced. 
Thus, $c_{\rm v}(T)/T$ becomes sensitive to the temperature at the very low temperature region, which is observed in various heavy fermion materials.

At the low-temperature region ($T<25$ K), the suppression of $c_{\rm v}(T)/T$ of YbAl$_3$ is observed with an increase of temperature from $T=0$. 
The coherence temperature ($T_{\rm coh}\sim 35$ K) is one of the typically characteristic energy scale of YbAl$_3$. 
Thus, the correlation effect is also reduced by temperature at $T \sim T_{\rm coh}$, so that the decrease of $c_{\rm v}(T)/T$ from $T=0$ may occur.

In the present calculation, although we cannot confirm the temperature dependence of $c_{\rm v}(T)/T$ accurately at $T \rightarrow 0$ due to the numerical difficulties, we show the following possibility; $c_{\rm v}(T)/T$ increases (decreases) from $T=0$ for the non-interacting (interacting) case at the very low temperature region. 

On the other hand, the temperature dependence of $c_{\rm v}(T)/T$ has common property for any values of $\tilde{U}$, in which the peak structures appear at $T=T_{\rm peak}$. 
With further increase of temperature, $c_{\rm v}(T)/T$ turns to decrease beyond $T_{\rm peak}$ $e.g.$, $T_{\rm peak}\sim 0.0025$ Ry for $\tilde{U}=0$ and $T_{\rm peak}\sim 0.0007$ Ry for $\tilde{U}=0.015$ Ry, respectively. 

In the present case, the large DOS exists 
below the Fermi level. Here, the center of gravity of the peak position in the DOS is defined as $\omega \sim E_{\rm peak}$. When the energy scale of the temperature greatly exceeds the difference between $E_{\rm peak}$ and $E_{\rm F}$, $c_{\rm v}(T)/T$ decays monotonically as temperature. 
It is well known that Fermi distribution function is broadened due to temperature and the width roughly corresponds to $4T$. 
The change of gradient of the specific heat $c_{\rm v}(T)$ as a function of $T$ occurs at $T\sim (E_{\rm F}-E_{\rm peak})/4$, which is shown in Fig. \ref{dos} and the inset of \ref{cv}. For example, each center of gravity of the peak in the DOS is roughly located at $E_{\rm peak} = -0.015 \sim -0.01$ Ry for $\tilde{U}=0$ and at $E_{\rm peak} = -0.006 \sim -0.004$ Ry for $\tilde{U}=0.015$ Ry (shown in Fig. \ref{dos}). 
Then the gradients of the specific heats are changed at $T=0.0025 \sim 0.0035$ Ry and at $T\sim 0.0008$ Ry, respectively. Furthermore the peak temperature of $c_{\rm v}(T)/T$ is rather shifted to lower temperature region because of the effect of factor $1/T$ in comparison with the result of specific heat. 

We stress that this peak structure of $c_{\rm v}(T)/T$ results from the band structure (or the DOS structure) and the peak position is located at less than $1/4$ of the difference between $E_{\rm peak}$ and $E_{\rm F}$. 
With an increase of the interaction, the effective 4f $J=7/2$ level is shifted to the Fermi level, so that the peak position of $c_{\rm v}(T)/T$ is also shifted to lower temperature region.

The peak temperature of $c_{\rm v}(T)/T$ is observed at 70 K, the amplitude is 80 mJ$\cdot$K$^{-2}\cdot$mol$^{-1}$ and the Sommerfeld coefficient is 40 mJ$\cdot$K$^{-2}\cdot$mol$^{-1}$ in the experimental result. In the present calculation, for $\tilde{U} \sim 0.015-0.018$ Ry, the peak temperature of $c_{\rm v}(T)/T$ is located at $T_{\rm peak} \sim 0.0002 - 0.0006$ Ry, which roughly corresponds to 30 $-$ 90 K. 
The amplitude of $c_{\rm v}(T)/T$ is range from 40 to 60 mJ$\cdot$K$^{-2}\cdot$mol$^{-1}$. 
For the value of $\tilde{U}$ to obtain this result of $c_{\rm v}(T)/T$, the Sommerfeld coefficient becomes 20 $-$ 30 mJ$\cdot$K$^{-2}\cdot$mol$^{-1}$ at the very low-temperature region (shown in Figs. \ref{gamma} and \ref{cv}).
Although there are the minor disagreements, these obtained results can reproduce the experimental ones for YbAl$_3$ semi-qualitatively. 

\section{Summary and discussions}
We have constructed the low-energy effective Hamiltonian with the realistic band structure and discussed quantitatively the specific heat 
for typical heavy fermion compound YbAl$_3$. 

The band term consists of the conduction electrons with NEF method, the localized 4f electrons and the hybridization term between the conduction and the 4f electrons which still keeps the angular dependence of the wave-vector $\mathbf k$. This model includes only a few parameters. 

After diagonalizing the band term of the Hamiltonian, we extract 14 bands near the Fermi level, which reproduces the LDA result. 
We further construct the low-energy effective Hamiltonian in order to consider the correlation effect which is mainly caused by the 4f $J=7/2$ electrons. 
By means of the self-consistent second-order perturbation theory with respect to the Coulomb interaction combined with local approximation, the DOS, the Sommerfeld coefficient and the specific heat divided by temperature are calculated. 

The Sommerfeld coefficient and $\lim_{T \rightarrow 0}c_{\rm v}(T)/T$ are also enhanced with an increase of the effective Coulomb interaction. 
The ratio $c_{\rm v}(T)/T$ decreases with an increase of temperature for $\tilde{U}>0$ from $T=0$, but the peak structures appear at higher temperature for any $\tilde{U}$. 

Thus, we stress that the characteristic structures of $c_{\rm v}(T)/T$ in YbAl$_3$ may originate from the correlation effect and the structure of the non-interacting density of states, respectively. 
Although some disagreements are left, the results for $\tilde{U}\sim 0.015 \sim 0.018$ Ry are in agreement with the experiments. 

The peak structure is experimentally observed at $\hbar \omega \sim 0.25$ eV in the optical conductivity, whose main contribution comes from the effective 4f $J=7/2$ level. 
While the peak structure of $c_{\rm v}(T)/T$ mainly is also caused by the effective 4f $J=7/2$ level, the corresponding peak position becomes rather smaller than the energy difference between the effective 4f $J=7/2$ level and the Fermi level. 
We can roughly understand the specific heat and the optical conductivity on equal footing within the present effective Hamiltonian. 

In conclusion, the good agreement with the experimental results reflects the appropriate description of the band structure in our model. Thus, our effective Hamiltonian succeeds in describing the realistic heavy-fermion materials with conduction bands mainly composed of s- and p- electrons.

\section*{Acknowledgement}
The authors would like to thank H. Harima and H. Okamura for valuable discussions. 
This work was supported by Grant-in-Aid for Young Scientists (B) (No. 19740195) from the Ministry of Education, Culture, Sports, Science and Technology of Japan. 

\end{document}